\documentclass[11pt]{article}
 \usepackage{amssymb}
 \def\Z{\mathbb{Z}}

 \def\pa{\partial}
 \def\k{\kappa}

 \def\b{\beta}
  
 \def\e{\epsilon}
 
 \def\k{\kappa}
  \def\L{\Lambda}
 \def\m{\mu}

  \def\O{\Omega}
 
 \def\be{\begin{equation}}
 \def\ee{\end{equation}}
 
\begin{document}

 \begin{flushright} BRX TH-426\\
       DFF 398/12/02 \\
       UPRF-2002-14
 \end{flushright}

 \begin{center}
{\Large\bf Large Gauge Invariance in Nonabelian  Finite
Temperature  Effective Actions}

S. Deser\\ Department of Physics, Brandeis University, Waltham, MA
 02454, USA\\ L. Griguolo\\ Dipartimento di Fisica, Universit\'a di Parma,
 INFN, Gruppo Collegato di Parma,\\ Parco Area Viale delle Scienze 7/A, 43100
 Parma, Italy
 \\ D. Seminara\\ Dipartimento di Fisica, Universit\'a di Firenze,
 INFN, Sezione di Firenze,\\
 Via G.\ Sansone 1, 55125, Sesto Fiorentino, Florence, Italy.
 \end{center}

 \begin{abstract}

We analyze large gauge invariance in combined nonabelian and thermal QFT
and their physical consequences for $D=3$  effective actions. After briefly
reviewing the structure of bundles
and large gauge transformations that arise in non-simply connected 3-manifolds
and gauge groups, we discuss their connections to Chern-Simons terms and
Wilson-Polyakov loops. We then provide an invariant characterization of
the ``abelian'' fluxes encountered in explicit computations of finite
temperature effective actions. In particular we relate, and provide explicit
realizations of, these fluxes to a topological index that measures the
obstruction to global diagonalization of the loops around compactified time.
We also explore the fate of, and exhibit some everywhere smooth, large
transformations for non-vanishing index in various topologies.

\end{abstract}

 \section{Introduction}
\setcounter{equation}{0}
Finite temperature gauge physics differs significantly from its
non-thermal counterpart.  The geometrical features related to the
presence of $S^1$-compactified Euclidean time considerably
complicate the zero-temperature picture. Already for abelian
effective actions (Dirac determinants) generated by integrating
out charged fermions in D=3, a number of unexpected phenomena
emerge \cite{DGS1,DGS3} through the interplay between ``large"
gauge invariance requirements and the topological properties of
non-simply connected spacetimes. Through careful regularization,
it was possible to deal with the usual array of ``global"
properties, such as spectral asymmetries (characteristic of odd
dimension), parity anomalies, preservation of large gauge
invariance and the role of (generalized) Chern--Simons (CS) terms.
In the process, we were able to exhibit very general ``Fourier"
representations of these $U(1)$ actions such that the large gauge
transformations (LGT) arising at finite temperature were properly
expressed with respect to the corresponding, non-perturbative,
Ward identities. In turn, this led to gauge-invariant expansions
of the effective actions useful for both large and small fermion
masses, where  the anomalies and ambiguities could also be
analyzed.

The object of the present work is the natural but non-trivial extension
of the above analysis to nonabelian theories, for various -- simply
connected or not -- internal symmetry groups such as $SU(2)$ and $SO(3)$.
The basic issue remains that of large gauge invariance, but with the added
nonabelian gauge complication. It is known that (as in the abelian case)
massive
fermions in a background nonabelian gauge field at $T\neq 0$ induce, at
one-loop level, a CS term whose coupling parameter is continuous in
temperature \cite{NS,Pisarski,NS2}. Therefore, at generic $T$, this parameter
cannot have the discrete value required for invariance under LGT.
On the other hand we could expect \cite{DGS1,Dunne,DasDunne}, from
the abelian analysis, that at
finite temperature an infinite number of terms is induced in the
effective action in such a way that large gauge invariance is
restored, even though it would not be manifest at any finite
perturbative order. The above statement can be easily proved by
studying the relevant Dirac determinant using $\zeta$-function
regularization, a manifestly gauge-invariant method providing a
well-defined geometrical tool for discussing global properties.
While the spacetime aspects of the Dirac determinants (now
augmented by the color index entries) in this formalism are very
similar to the abelian case, so the various series expansions
presented in \cite{DGS1} carry over formally unchanged, this
procedure unfortunately gives insight neither into the mechanism
through which the perturbative series organizes itself nor into
the nonperturbative information, such as holonomies, indices and
fluxes, needed to explain the global invariance. Indeed, we shall
see that significant non-perturbative changes arise precisely for
configurations that involve a novel index.

The feasibility of an explicit description of the large gauge
invariance properties in the abelian case is related to the
peculiar structure of the $U(1)$ gauge group.
Abelian gauge transformations with non-trivial winding number
exist only at finite temperature, where the CS action jumps
discontinuously by a non-vanishing boundary term when magnetic
flux is present on the underlying two-dimensional manifold.
Conversely, it was possible to show that the only degree of
freedom of the gauge field transforming significantly under LGT
was the flat connection wrapping around $S^1$. Large gauge
invariance is precisely the statement that the effective action
must be periodic in the flat connection, implying in this way an
additional finite Ward identity. Moreover, this critical,
topological, degree of freedom  affects the parity violating (PV)
of the action only when accompanied by magnetic flux on the
corresponding two-dimensional manifold. In the abelian case,
therefore, the non-trivial effects of LGT are unavoidably linked
to the simultaneous presence of a magnetic flux and of a flat
connection.\footnote{The above picture, developed in \cite{DGS1},
is itself a natural extension of one in zero space dimension
\cite{Dunne}, the important new feature being the magnetic flux.}
Our general considerations were illustrated by explicit
integration in presence of some specific, physically non-trivial,
field configurations\footnote{Our computations were mainly
performed with vanishing electric field $\mbox{\boldmath $E$}$
purely in order to permit analytical results, since
$\mbox{\boldmath $E$}$ plays no role in preserving large gauge
invariance: Indeed, later constructions with $\mbox{\boldmath $E$}\neq 0$
 using improved derivative
expansions \cite{Salcedo} or partially resummed perturbation theory
\cite{bdf} confirmed our conclusions.} \cite{DGS1} (see also
\cite{Fosco:1997ei}).
While the topological complications of abelian theory arose in a very
specific and ``rigid'' way, disappearing in the (topologically trivial)
limit of zero temperature, nonabelian LGT and some of their consequences
persist at $T=0$,
%
as illustrated by the well-known quantization \cite{zeroT} of the
tree-level CS coefficient. At one loop and zero temperature,  the non-trivial
topological part of the PV action is entirely governed by a gauge
invariant functional of the field, usually
denoted as $\eta(0)$, that is neither local nor continuous. It can
be explicitly computed with the help of the Atiyah-Patodi-Singer
theorem \cite{aps} and consists of two parts: a continuous local
functional given by the CS action plus a nonlocal, discontinuous,
contribution given by a topological index\cite{aps1}. Large gauge invariance
is maintained through a cancellation between the two. While,
for massless fermions, this is the only PV-contribution,
for massive fermions, new (but harmless) structures can be
generated: their form in the limit of small and large mass is
discussed in \cite{DGS1}.
A direct zero-temperature perturbative computation of the PV effective
action produces a properly normalized CS term that respects large gauge
invariance in the familiar sense: the quantized nature of the CS coefficient
is preserved.

For massive fermions, when the
temperature is turned on, however, the very same expansions start,
discontinuously, to generate $T$-dependent CS coefficients exactly
as in the abelian case, requiring a combined, thermal + gauge,
description.  In the easiest case, when the gauge group is, like
$SU(N)$, simply connected, the basic topological properties are
unaltered; at the same time the presence of a non-trivial $S^1$
and the explicit form of the perturbative results suggests that
the zero-mode of the time component $A_0$ will again play a
special role. To be more precise, the (untraced) Wilson-Polyakov
loop (WPL) around Euclidean time is the important new ingredient,
just as it was in the abelian case. Unfortunately, in the nonabelian case,
flat connections do not behave simply under LGT and it is not
immediately obvious what kind of fluxes could be coupled there,
since as we will review (in the $SU(N)$ case) the bundle structure
is trivial. At the same time, the explicit example of \cite{RSF},
the perturbative computations of \cite{Das} and the derivative
expansions of \cite{Salcedo} suggest that, at least for vanishing
$\mbox{\boldmath$E$}$, the effective actions bear a strong
resemblance to the abelian ones, requiring also a field endowed
with an apparent magnetic flux. We will start our analysis by
examining to what extent the thermal abelian mechanism can be
embedded in the more complicated  nonabelian setting. In the process, we will encounter a new (in this
context) topological index, associated with the obstruction to
diagonalizing WPL globally, and playing the role of a generalized
magnetic flux\footnote{This index has already appeared in $QCD_4$ in the
discussion of the abelian projection and of the related 't Hooft-Polyakov
monopoles \cite{Wipf,QCDabelian,QCD42}.  It was, in fact, used to give a
topological gauge invariant characterization of this family of solutions.
In three dimensions, its role has been also discussed by \cite{BT}.}.
Moreover, we shall find that a  class of LGT
can be studied explicitly, for specific field configurations,
within a particular gauge fixing; the new topological index is
also pivotal to this analysis. Although our embedding is
completely general, only in the case of vanishing
$\mbox{\boldmath$E$}$ does it provide an exhaustive account of the
gauge invariance problem: for generic field configurations, it
must be accompanied by other mechanisms to deal with the non-flat
part of $A_0$, {\it i.e.} the electric sector. We believe, however,
that the LGT properties of the
magnetic sector are always controlled in the above manner.

Our paper is arranged as follows: in Sec.\ 2, we set out some
basic facts about the topology of internal gauge groups, mappings
into spaces of the form $S^1 \times$ compact Riemann surfaces of
arbitrary genus, the structure of the CS term and its finite
temperature transformation properties. In particular, we try to
extend our earlier, purely 3-dimensional, abelian analysis of the
CS coefficient's  quantization at finite $T$ to the nonabelian
$SU(N)/\Z_N$ without going through a 4-dimensional embedding argument
\cite{DW}. We show how, for these non-simply connected groups,
the zero-temperature condition given by the
third homotopy group may be modified due to the presence of a
non-trivial first homotopy group. In the process, we present and
motivate an explicit, simple, smooth and periodic LGT on
$S^1\times S^2$, followed later by one on $S^1\times T^2 \sim T^3$.
In Sec.\ 3, we turn our attention to WPL, whose analysis was
equally central in the abelian case, because it carries the
global gauge invariant information. We note that, while this
gauge invariant information is of course
contained in its eigenvalues, diagonalizing WPL is, in general,
not a smooth operation:  this quantity, viewed as a map from the base
manifold into the gauge group, might not be not globally diagonalizable.
Using the formalism of \cite{BT}, we associate an index to this topological
obstruction and show (by going to a suitable, ``diagonal", gauge)
that magnetic fluxes appear along the Cartan directions, despite
the vanishing of the original bundle's first Chern class. In Sec.\
4,  we explicitly construct the static temporal gauge $\dot A_0=0$
in order to analyze the structure of the potentials. We discover a
class of LGT preserving this gauge condition when the WPL has a
non-trivial index, and transforming the WPL holonomy exactly as
for the $U(1)$. Using the above ingredients in Sec.\ 5, we treat
configurations with (for simplicity) vanishing electric field
along the lines of \cite{DGS1}, the role of magnetic flux being
played here by an array of gauge-invariant magnetic projections
related to the above index.  A final section summarizes our
results and outlines some open problems.

\section{LGT and Topological Charges}
\setcounter{equation}{0}
\subsection{D=3 bundles}
Let us first review some basic facts  about Euclidean thermal
gauge fields on 3-spaces of the form ${\cal M}=S^1 \times \Sigma_g$,
where $\Sigma_g$ a compact Riemann surface of genus $g$.  The
circle $S^1$, parameterized by the Euclidean time, $T$, has
circumference  $\b = 1/\k T$.

In the following, we shall focus on principal $G$-bundles having
${\cal M}$ as basis and $G$ as structure group; their topological
classification is more involved than for $T$=0. There, one usually
compactifies Euclidean three-space on $S^3$, the three-sphere. Now
principal $G$-bundles over $S^3$ are classified by $\pi_{2}(G)$
\cite{ste}, the second homotopy group of $G$; however $\pi_2(G)=0$
for any Lie group and hence, in the usual $S^3$ picture, the
bundle associated with the gauge fields is automatically trivial.
The situation changes for more general three-manifolds ${\cal M}$,
since $G$-bundles are now classified by elements of the cohomology
group $H^2({\cal M},\pi_1(G))$ \cite{ais}. We see that non-trivial
topological charges could arise when non-simply connected groups
are taken into account; the commonly considered $SU(N)$ case is,
instead, still trivial, since $\pi_1(SU(N))=0$. If $G=U(N)$ then
$\pi_1(G)=\Z$ and the bundle is classified by an element of
$H^2({\cal M},\Z)$, which may be identified with the first Chern
class.  At finite temperatures, where ${\cal M}=S^1\times\Sigma_g$ is
the relevant base manifold, we find
\begin{equation}
H^2(S^1\times\Sigma_g,\Z)={\Z}^{2g+1}\; ,
\end{equation}
upon using the explicit form of the first Betti number of ${\cal M}$,
$b_1({\cal M})=2g+1$, and the absence of torsion \cite{is,vick}. This
clearly generalizes the situation of $N=1$ and $g=0$ (the 2-sphere),
by allowing for (several) magnetic fluxes.
[Gauge groups other than $SU(N)$ may also
effectively arise in simple situations, for example when adjoint
fermions are coupled to $SU(N)$ gauge fields.]
Another interesting
case is $G=SO(N)$; then $\pi_1(G)=\Z_2$ and the bundle is
classified using $H^2({\cal M},\Z_2)$, which may be identified
with the second Stiefel-Whitney class. Using Poincar\'{e} duality
and some basic properties of homology theory (see for example
\cite{vick}), one can compute
\begin{equation}
H^2(S^1\times\Sigma_g,\Z_2)={\Z_2}^{2g+1}.
\end{equation}
A third simple example is given by taking $G=SU(N)/\Z_N$: in this
case one can show along the same lines that
\begin{equation}
H^2(S^1\times\Sigma_g,\Z_N)={\Z_N}^{2g+1}.
\label{zenne}
\end{equation}

The above classifications are interesting because they exhaust the
possible topological sectors of finite temperature D=3, providing
the analogue of the instanton charges for the gauge groups
considered. They will help us understand the permissible values of
the CS coefficient and the transformation properties of the CS
action itself.
 At finite
temperature, one must specify the boundary conditions on $S^1$:
the gauge connections can always be chosen periodic,
\begin{equation}
A_\mu(t+\beta,  \mbox{\boldmath$x$})=A_\mu(t,\mbox{\boldmath$x$}
).
\end{equation}
Fermions can be taken to be antiperiodic,
 \begin{equation} \label{abebound1}%
 \psi(t+\beta ,  \mbox{\boldmath$x$})=-\psi(t,\mbox{\boldmath$x$})
 \end{equation}
if they are in the fundamental representation, or antiperiodic up
to a global $\Z_n$ transformation ({\it i.e.}, up to an
element of the center), if in the adjoint representation.
Therefore the effective gauge group is $SU(N)/\Z_N$, which
allows non-trivial topological sectors.

\subsection{LGT on non-simply connected manifolds and groups}
We have seen that the presence of a non-trivial $\pi_1({\cal M})$
can change topological structure;  the natural problem now is to
understand how LGT are changed when the base manifold (and
possibly the gauge group) is not simply-connected. For general
${\cal M}$ and $G$, under the simplifying assumption of dealing
with trivial bundles\footnote{For a non-trivial bundle $E$ the classification
of LGT is equivalent to that of Map$_G(E,Ad(G))$ (see \cite{Husemoller}).
Fortunately, we do not need to study this more complicated question.},
 LGT  are classified by the set of the homotopy
classes [${\cal M},G$] of continuous maps from ${\cal M}$ into
$G$. In the zero-temperature case, when ${\cal M}$ is $S^3$ we
simply have [${\cal M},G$]=$\pi_3(G)$, leading to the usual
classification in terms of integer winding numbers \cite{JR}. For
$SU(N)$, we know that $\pi_3(SU(N)) =\Z$. The general case is more
involved: the homotopy classes [${\cal M},G$] are labelled by two
topological winding numbers $(W_1, W_2)$. The primary, $W_1$, and
the secondary, $W_2$, are elements of the following cohomology groups,
 \be
 W_1\in H^{1}({\cal
M},\pi_1(G))\;, \;\;\;\; W_2\in H^{3}({\cal M},\pi_3(G))\; \; .
 \ee
These topological charges are the obstructions to deforming a
generic map $U$ into a constant one. While this finer
classification is not really essential for $SU(N)$, whose $\pi_1$
vanishes, the primary winding number does become relevant for
non-simply connected groups like $U(N)$ or $SO(N)$. Moreover, for
any compact and orientable ${\cal M}$ it can be shown \cite{is}
that
\begin{equation}
H^3({\cal M},\Z)=\Z\;  ,
\end{equation}
so the structure of  LGT for $SU(N)$ is basically unchanged at
finite temperature.

We next display the promised LGT example in $S^1 \times S^2$, and
(for simplicity) $SU(2)$:
 \begin{eqnarray} \label{bibo}%
 U_n(t,\theta,\phi)& \equiv &u(t) \hat U_n(t,\theta,\phi) \equiv
 \exp\left (-i~\frac{\pi}{\beta}~t~\sigma_3\right) \exp\left
(i~\frac{\pi}{\beta}~t~ \mbox{\boldmath $\omega$}_n \cdot
\mbox{\boldmath $\sigma$} \right)
 \nonumber\\
  & = &\exp\left (-i~\frac{\pi}{\beta}~t~\sigma_3\right)
 \left [\cos\left(\frac{\pi}{\beta}~t\right)I\!\!I+i
 \sin\left(\frac{\pi}{\beta}~t\right) \mbox{\boldmath $\omega$}_n
 \cdot \mbox{\boldmath $\sigma$} \right] \; .
 \end{eqnarray}%
In terms of the two-sphere angles $(\theta , \phi )$, the
components of the unit 3-vector $\mbox{\boldmath $\omega $}_n$ are
$\omega^1_n(\theta,\phi)=\sin\theta\cos( n\phi)$,
$\omega^2_n(\theta,\phi)=\sin\theta\sin( n\phi)$ and
$\omega^3_n(\theta,\phi)=\cos\theta$. A similar
transformation\footnote{Any other large transformation
$\bar{U}_n$ will differ from (2.8) by a small one. One recent
example is that of \cite{DAT}, which was obtained from the usual
$T=0$ form \cite{zeroT}, $\bar{U}_n^L (T=0) = \exp [n\pi i\,
\mbox{\boldmath $y$} \cdot \mbox{\boldmath
$\sigma$}/\sqrt{\mbox{\boldmath $y$}^2 + \lambda^2}\, ]$, by
compactifying time through the mapping $t\rightarrow \tau \equiv
\tan \: \frac{\pi t}{2\beta}$. Here $\mbox{\boldmath $y$} \equiv
(t, \mbox{\boldmath $x$}  )$ and $\lambda$ is an arbitrary
constant cutoff. However, as noted in \cite{DAT}, the price of
this mapping is that (in contrast with (\ref{bibo})) the resulting
$\bar{U}^L_n$ is not analytic.} was obtained, in a different
context, in \cite{Percs}; there it was noted that $\hat{U}_n$
alone was antiperiodic, but that it could be made periodic simply
by composing it with the ``small" antiperiodic $u(t)$ that
(being angle-independent) cannot alter the winding in $\hat{U}_n$. This
$\hat{U}_n$ is just a square root of the well-known periodic
ansatz later introduced independently in \cite{Pisarski}, which is
why it produces all windings, but at the price of antiperiodicity.
[The original, periodic, ansatz covered only even windings.]  A
simple computation confirms that the contribution from $u(t)$ to
the winding integrand is a total divergence whose integral
vanishes on the sphere. We notice that for $n=0$ the representative is
not the identity but a particular, non-trivial, small gauge transformation.
It is possible to generalize the above
construction to $SU(N)$ by embedding the $SU(2)$ example into $SU(N)$:
 the antiperiodicity of
$\hat{U}$ is repliced by  ``periodicity" up to an element of the center
$\Z_N$ and $u(t)$ by a similar small
transformation. In Sec.\ 3 we will present an
example with different topology, the torus $T^3$.

Returning to the classification problem, let us first contrast the
above treatment of $SU(N)$ with that of $U(1)$: For the latter, only $W_1$ is
relevant ($\pi_1(U(1))=\Z$), because its $\pi_3$ vanishes. [At
finite temperature, abelian LGT  therefore appear, whose effects
on the quantization of the CS coupling constant and on the
definition of the CS action have already been mentioned.] Here we
simply write the result for generic ${\cal M}$,  easily
derivable from \cite{is}:
\begin{equation}
H^1({\cal M},\Z)=(\Z)^{2g+1}.
\end{equation}
The gauge groups $U(N>1)$ and $SO(N>3)$, instead,
experience the combined effect of both winding numbers:
 \be
\left[{\cal M},U(N)\right]= W_1\oplus W_2=(\Z)^{2g+1}\oplus \Z \;, \;\;\;
\left[{\cal M},SO(N)\right]=W_1\oplus W_2=(\Z_2)^{2g+1}\oplus \Z \; .
 \ee
The case of $SO(3)$ is special: not all possible primary and
secondary winding numbers separately label LGT, but only those
having a particular relation. In \cite{is} it was shown that $W_1$
and $W_2$ must satisfy $(W_1)^3=W_2\,{\rm mod}\,2$ as a relation
in $H^3({\cal {\cal M}},\Z_2)$: For  ${\cal M} = S^1\times \Sigma_g$, it is not
difficult to prove that the above condition can be satisfied
trivially since $(W_1)^3=0$ and one gets
\begin{equation}
[{\cal M},SO(3)]=(\Z_2)^{2g+1}\oplus 2\Z,
\end{equation}
where the second term on the r.h.s. simply realizes the condition
$W_2\,{\rm mod}\,2=0$: only even secondary winding numbers appear
in the classification of  LGT for $SO(3)$.

\subsection{Quantization of CS coefficient on non simply-connected
groups and manifolds}
While the systematic analysis of D=3 LGT \cite{is} was originally
motivated by studying the vacuum structure of four-dimensional
gauge theories, their relevance to $D$=3 physics was realized by
analysis of CS theory \cite{zeroT}, which we review briefly. When
defined on a trivial $G$-bundle over ${\cal M}$, the CS action has
the form
\begin{equation}
S=2\pi k I_{CS} = \frac{k}{4\pi}\int_{{\cal M}} d^3x \,\epsilon^{\alpha\mu
\nu} ~{\rm Tr}\left[ A_\alpha
\partial_\mu A_\nu - \frac{2i}{3} A_\alpha A_\mu A_\nu\right] \;.
\label{trics}
\end{equation}
The behavior of CS under gauge transformation,
\begin{equation}
A^U_\mu=U^{-1}A_\mu U+iU^{-1}\partial_\mu U \; ,
\end{equation}
is given by
\begin{equation}
\label{derri}
I_{CS}[A^{U}]=I_{CS }[A]+{\cal W}[U]-\frac {i}{8\pi^2}\int d^3
x\,\epsilon^{\alpha \mu \nu}\partial_\alpha {\rm Tr}[A_\mu U^{-1}
\partial_{\nu} U],
\end{equation}
with
\begin{equation}
\label{W3}
{\cal W}[U]=\frac{1}{24\pi^2}\int_{{\cal
M}}d^3x\,\epsilon^{\alpha\mu\nu} {\rm Tr}\left[
U^{-1}\partial_{\alpha}U U^{-1}\partial_{\mu}U U^{-1}
\partial_{\nu}U\right].
\end{equation}
Since the bundle structure is trivial, the total derivative term
can be neglected (our manifolds having no boundary); for $SU(N)$,
${\cal W}[U]=W_2[U]$ measures the secondary winding number or topological
degree of $U$. This in turn leads to the integer quantization
condition on the coupling constant $k$, since the phase, $S$,
is the relevant physical object at quantum level.
Of course when $\pi_1(G)\neq 0$, the situation may
change because the primary winding number can also enter for
non-trivial $G$-bundles. In particular, the CS term itself is not
well-defined when the bundle is non-trivial: the $A_\m$ in
(\ref{trics}) are not globally defined as one-forms on ${\cal M}$
and the action is patch-dependent by virtue of (\ref{derri}).
[Here the total derivative contributions must also be taken into
account.] The simplest, but physically relevant, realization of
this possibility was actually in finite temperature abelian theory
\cite{DGS1,DGS3,poly,Jackiw:1997kp}, where the non-trivial bundle
structure forced an
analogous quantization requirement. The total derivative term in
(\ref{derri}) contributes when LGT (necessarily of primary type)
are performed in presence of topological charges (the analog of
instantons).  As a result, we noted that, for the CS action
 to be well-defined, a doubling of the naive
quantization condition was required. A general approach to the
problem was proposed in \cite{DW} by defining the CS action as
\begin{equation}
S=\frac{k}{4\pi}\int_{{\cal B}}d^4x\,\epsilon^{\alpha\beta\mu\nu}{\rm Tr}
\left[F_{\alpha\beta}F_{\mu\nu}\right],
\end{equation}
where ${\cal B}$ is some 4-manifold (or more generally 4-complex)
whose boundary is ${\cal M}$
and the $G$-bundle has been extended over ${\cal B}$. Now it is
the  ${\cal B}$-independence of $S$ that forces
$k$ to be an integer, but demanding that the $G$-bundle be
extendable poses further restrictions on $k$. We argue
here that these restrictions, consistent with our quantization
rule in the abelian case, can also be understood as coming from
LGT within the purely three-dimensional approach, at least for
${\cal M}=S^1\times S^2$. To show this, consider $SU(N)/\Z_N$: we have
shown above that non-trivial bundles can be present, while
existence of LGT can also be easily inferred, since now the
$SU(N)$ gauge functions need be periodic only up to a $\Z_N$
transformation. A particularly explicit realization can be offered
for $SO(3)$, equivalent to $SU(2)/\Z_2$, using the adjoint
representation of $SU(2)$, with generators
$(\tau_a)_{bc}=i\epsilon_{abc}$. LGT on ${\cal M}$ are labeled by
$\Z_2\oplus 2\Z$ and are explicitly realized by \cite{Percs}
\begin{eqnarray}
\label{bibo1} U_{0,2n}(t,\theta,\phi) & = &\exp\left
(-i~\frac{2\pi}{\beta}~t~\tau_3\right) \exp\left
(i~\frac{2\pi}{\beta}~t~ \mbox{\boldmath $\omega$}_n \cdot \tau
\right) \nonumber\\
U_{1,2n}(t,\theta,\phi)&=&\exp\left (i~\frac{2\pi}{\beta}~t~
\mbox{\boldmath $\omega$}_n \cdot \tau \right)\; .
\end{eqnarray}
We must remark that the $U(1)$ factor in the first of these two classes
of transformations is actually a large transformation by itself: it
cannot be contracted continuously to unity; only its square can.
Its primary winding number is, indeed, $1$. Since the second factor
also has primary winding number $1$, their product possesses a vanishing
$W_1$. Recall that $W_1$ takes values in $\Z_2$.

\noindent
Therefore, the non-trivial $\Z_2$ classes are related to a non-trivial loop
around $S^1$, similar to the  LGT of $U(1)$: at $n=0$ it is more
convenient to use a simpler representative of the non-trivial class,
depending only on $t$
\begin{equation}
{\cal U}_{1}(t)=\exp\left(-i~\frac{2\pi}{\beta}~t~\tau_3\right).
\end{equation}
In the following we will work (for general $N$) in the fundamental
representation, realizing the quotient at the level of gauge
functions, periodic up to the particular element of $Z_N$ that is
identified with the primary winding number. The basic
transformation is
\begin{equation}
\label{repre}
{\cal U}_{1}(t)=\exp\left(-i~\frac{2\pi}{N\beta}~t~H\right),
\end{equation}
where $H$ is the diagonal, traceless, $N\times N$ matrix with
entries $(1,1,..,1,1-N)$; the elements of the other classes are
obtained by taking all powers up to $N$--1. We have now
constructed the representatives for primary windings number (at
genus zero the classes are in correspondence with $\Z_N$). Next we
 construct the non-trivial $SU(N)/\Z_N$ bundles on
$S^1\times S^2$: as in the $U(1)$ case; one can think of them as
coming from monopole-like configurations on the sphere.
Non-trivial connections on $S^2$ are characterized by a
non-trivial $\Z_N$ holonomy around, say, the north pole: a simple
way to obtain this is to fix the transition functions, labelled by
the various elements of $\Z_N$, and taken at $\theta=\pi/2$ for
definiteness, to be
\begin{equation}
G_{m}(\phi)=\exp\left(-i~\frac{2\pi m}{N\beta}~\phi~H\right)\; ,
\;\;\;\; m=0,1..N-1 \;  .
\end{equation}
 We promote $G_m$ to be the transition function of
the full $SU(N)/\Z_N$-principal bundle over $S^1\times S^2$ ($m$
being the number associated to the different bundle structures,
according to eq.(\ref{zenne})). From the three-dimensional point
of view $G_m$ is now  seen to be the transition function at the
intersection of two patches covering ${\cal M}$, having toroidal topology
(we have implicitly assumed that our transition functions could be
chosen time-independent). To construct the CS action in this
non-trivial case, we can resort to a patch by patch definition:
\begin{equation}
\label{patch}
I_{CS}(A)=I^{X_1}_{CS}(A_1)+I^{X_2}_{CS}(A_2),
\end{equation}
where $X_1,X_2$ are two solid tori with oppositely oriented
boundaries $T^2$, \, $A_1,A_2$ being the expressions for the gauge
connections (now globally defined on $X_1,X_2$) with boundary
values related by the transition function $G_{m}$. As in the
abelian case, the definition (\ref{patch}) must be augmented by a
term depending explicitly on the transition function, in order to
be independent of the particular local trivialization of the
bundle (see \cite{alvo,poly} for the cohomological meaning of
these terms). Our proposed generalization
of (\ref{patch}) is
\begin{eqnarray}
\label{patch2}
I_{CS}(A)&=&I^{X_1}_{CS}(A_1)+I^{X_2}_{CS}(A_2)+\hat{\cal W}[G]\nonumber\\
&+&\frac {i}{8\pi^2}\int_{T^2}d^2x\,\epsilon^{ij}{\rm Tr}[G\partial_{i}G^{-1}A_{1j}],
\end{eqnarray}
where $\hat{W}[G]$ is the WZWN action associated to $G$. The origin
of (\ref{patch2}) can be understood by recalling eq. (\ref{derri}).
It can be
checked that for trivial bundles it coincides with the CS action,
when written in term of globally defined connections,  and that it
reproduces the abelian result. The definition of
the WZWN action implies that the CS coupling parameter $k$ must be
even for $N$ even,  and an arbitrary integer for $N$ odd,
according to the general analysis for $SU(N)/\Z_N$ presented in
\cite{fgk}. Gauge invariance further restricts $k$: let us compute
the gauge variation of (\ref{patch2}) under
${({\cal U}_{1})}^s\equiv{\cal U}_{s}$ ($s=1,2..N-1$), the LGT with
non-vanishing primary winding number, represented as in (\ref{repre}).
Since ${\cal U}_s$
commute with the transition function $G_{m}$, they are
globally defined automorphisms of the bundle. We obtain
\begin{equation}
\delta I_{CS}=\frac {1}{8\pi^2}\int_{T^2}d^2x\,\epsilon^{ij}{\rm Tr}
[G_m\partial_{i}G_m^{-1}({\cal U}^{-1}_s\partial_j {\cal U}_{s}-{\cal U}_{s}\partial_j
{\cal U}_s^{-1})].
\end{equation}
Using the explicit form of $H$, this variation reduces to
\begin{equation}
\delta I_{CS}=ms\frac{N-1}{N}.
\end{equation}
From the extreme case,
$m=s=1$, we learn that $k$ must be a multiple of $N$ in order
for the quantum phase to be well-defined. In \cite{DW} the same
problem was examined, following the four-dimensional route and
using algebraic geometrical techniques; our more down-to-earth
approach can reproduce that result for $N$ odd, but for $N$ even
we apparently miss a factor 2 (the quantization rule appearing in
\cite{DW} is $k=2N$ for $N$ even). One possible explanation may be
that an extra factor $1/2$ is needed in the normalization of the
CS term for $N$ even: it is well-known \cite{fgk,gw} that WZWN
theory needs this extra factor in its definition, when quotiented
by $\Z_N$ for $N$ even. If the CS action experienced the same
change we would find full agreement with \cite{DW}; in any case,
the general $N$-dependence displayed by our approach indicates
that a purely three-dimensional definition is indeed possible.

\section{Fluxes and their Characterizations}
\setcounter{equation}{0}
\subsection{General Framework}
For finite $T$, understanding how the invariance of the effective
action under LGT is explicitly realized can be difficult. In the
abelian case, we singled out the flat connection (effectively WPL)
wrapping around $S^1$ as the carrier of this information, leading
to the basic requirement that the effective action be periodic,
{\it i.e.}, have ``Fourier" form in WPL.  A major role in that
analysis was also played by fields with non-vanishing Chern class.
[In fact, most of the examples in the abelian literature exploit
this property to generate a candidate CS term in the effective
action.] For $SU(N)$, the interplay between periodic time, LGT,
and fluxes appears to be weakened: unlike in the abelian case, LGT
survive in the zero-temperature limit and there are no bundles
with non-vanishing first Chern class. Therefore the generic
nonabelian picture cannot simply mimic the abelian one -- some
new mechanism must be introduced. Nevertheless, in the following,
we show that there is a vast class of nonabelian fields with
features common to those of the abelian analysis and for which one
can define apparently abelian, but fully gauge invariant, fluxes.

As stated, the additional abelian information carried by the flat
connection is encoded in the WPL about $S^1$. It is natural to
expect that the same quantity will also describe some topological
properties of the effective action, such as its behavior under
LGT. Here, as we shall see, the WPL is related to a new
topological number, whose role is very similar to that of magnetic
flux. To illustrate this unexpected connection, we must first
introduce some results about maps from a two-manifold into
$SU(N)$. In fact, from the mathematical point of view, the
(untraced) WPL defined by
\begin{equation}%
\Omega(\beta,\mbox{\boldmath$x$})=P\exp\left(-i\int^\beta_0
dt^\prime A_0(t^\prime,\mbox{\boldmath$x$})\right ) \equiv {\O}\;
,
 \end{equation}%
is a map from the space manifold, say $S^2$, into $SU(N)$. Despite
the fact that the $\pi_2$ of $SU(N)$ vanishes, we can associate to
the applications from $S^2$ into the group an integer-valued
topological index \cite{BT} that measures the obstruction to
diagonalization of $\Omega$ by globally defined transformations.
This integer will play the role of a generalized magnetic flux.
[This is not a novel idea:  it was indeed widely exploited in
$D=4$ to study the invariant meaning of magnetic
monopoles and magnetic defects in the gauge where $A_0$ is
time-independent and diagonal \cite{QCDabelian} and in particular to
discuss the fate of LGT in the Hamiltonian approach \cite{QCD42}
to $QCD_4$.]

Following \cite{BT}, we shall review the invariant
characterization of the gauge connections for which this index
does not vanish.\footnote{The relevance of this class of fields in
$D=3$ physics was already underlined  in \cite{BT}, where these
configurations were shown to be crucial to the path integral
solution of CS theory and to the derivation of the Verlinde formula.}
We confine our analysis to regular WPL, which are dense in the
space of all possible maps.  An element $g$ in a group $G$ is
regular if the set of elements commuting with $g$ has dimension of
the maximal torus $T^m$ of $G$. [In $SU(N)$, this statement is
equivalent to saying that the eigenvalues of $g$ are
non-degenerate.] Our problem may then be formulated as follows:
Can a given smooth regular map $g: \Sigma \to G$ be written as
 \begin{equation}
 g=h^{-1} s h
 \end{equation}
where $s:\Sigma\to T^m$ and $h:\Sigma\to G$ are smooth and
globally defined maps? The answer to this question cannot be
always affirmative, as the following classic example shows.
Consider the $SU(2)$ element
 \begin{equation}
 g \equiv i \hat{\mbox{\boldmath$r$}} \cdot
 \mbox{\boldmath$\sigma$}\; , \;\; \hat{\mbox{\boldmath$r$}}^2 = 1
 \; .
 \end{equation}
Since $\hat{\mbox{\boldmath$r$}}^2 = 1$ is also a sphere, we can
regard $g$ as an application between two $S^2$, the second one
being the equator of $SU(2)$. To any such application $g$, we can
associate a topological winding number given by
 \begin{equation}
 \label{index}
 n(g) = -\frac{1}{32\pi}\int_{S^{2}} {\rm Tr}(g[dg,dg]) \ .
 \end{equation}
Clearly $n(g)$ vanishes identically  if $g$ can be diagonalized
smoothly. On the other hand, $n(g)$ is equal to 1 for (3.20),
as a simple calculation in polar coordinates shows; thus we must
conclude that the above $g$ cannot be smoothly diagonalized.
However, when we consider $g$ as a matrix-valued field  defined on
a bundle, conjugation to the maximal torus, {\it i.e.}, its
diagonalization, acquires the meaning of performing a gauge
transformation. Limiting the admissible gauge transformations just
to the globally defined ones is too restrictive; the natural
geometrical requirement is (as for the gauge connections) rather
that they be definable chart by chart, respecting the cocycle
conditions on the overlaps of different charts.\footnote{From the
mathematical point of view we want to exploit the possibility of
representing a given fiber bundle through any of the coordinate
bundles belonging to its equivalence class. We call this -- by a
physicist's abuse of language -- a gauge transformation, even
though in mathematics that name is restricted to the vertical
automorphism of the bundle.}
In the following, we will diagonalize $g$ in this wider
arena, finding the new index in the process.

First, we must understand how to keep track of  the information
encoded in $n(g)\neq 0$ if we allow non-smooth transformations. To
this end, observe that the index $n$ can be made invariant also
under non-globally defined transformations if there is  a gauge
connection on $S^2$; this can be achieved by introducing an arbitrary
one to enlarge the space of
allowed transformations sufficiently. But  our physical framework
already possesses a natural candidate -- the spatial gauge
potential  $\mbox{\boldmath$A$}$. Then, according to general
theory, we can write
 \begin{equation}
 \label{genindex}
 n(g,A) = -\frac{1}{32\pi}\int_{S^{2}} {\rm Tr}g [d_{A}g,d_{A} g] -
 \frac{1}{2\pi}\int_{S^{2}}  {\rm Tr}[ gF_{A}] \, , \label{cno}
 \end{equation}
 with $d_{A}g=dg +[\mbox{\boldmath$A$},g]$ and $F_{A} =
 d\mbox{\boldmath$A$} + \frac{1}{2}[\mbox{\boldmath$A$},
 \mbox{\boldmath$A$}]$ is the associated magnetic field. The
 invariance under non-smooth transformation is manifest, since no
 integration by part is required to prove that (\ref{genindex}) is
 gauge invariant.  Rewriting  it as
 \begin{equation}
 n(g,\mbox{\boldmath$A$}) = -\frac{1}{32\pi}\int_{S^{2}} {\rm Tr}g [dg,dg] -
 \frac{1}{2\pi}\int_{S^{2}}  {\rm Tr} [d(g \mbox{\boldmath$A$})]
 \end{equation}
shows that it is equivalent to (\ref{index}) when $g$ and
$\mbox{\boldmath$A$}$ are smooth, since the second integral in
(3.23) vanishes.

 Let us now perform the $SU(2)$ transformation that diagonalizes
 $g=h^{-1}\sigma_3 h$; then
 \begin{equation}
 n(g,\mbox{\boldmath$A$})=1= -\frac{1}{2\pi}\int_{S^{2}} {\rm Tr}\left[ \sigma_{3}
 d(\mbox{\boldmath$A$}^{h^{-1}})\right].
 \end{equation}
In particular, if we introduce the abelian gauge connection
$\mbox{\boldmath$a$} \equiv -{\rm  Tr}\sigma_{3}
\mbox{\boldmath$A$}^{h^{-1}}$ we obtain
  \be n(g,\mbox{\boldmath$A$}) = 1 =
 \frac{1}{2\pi}\int_{S^{2}} d\mbox{\boldmath$a$}\; ;
 \ee
$\mbox{\boldmath$a$}$ can indeed be interpreted as a $U(1)$
connection, since its transition functions are abelian. The first
Chern class of the component of $\mbox{\boldmath$A$}^{h^{-1}}$
along the Cartan subalgebra (here $\sigma_3$ ), is equal to
the winding number of the original map. Geometrically, we have
picked up  a non-trivial torus bundle; physically, a non-vanishing
magnetic flux, related to the diagonal components of the field,
appears. We have thus associated to a $SU(2)$ connection a
topological number that plays a dual role: in  {\it regular}
gauges it measures the obstruction to diagonalizing WPL, while in
{\it singular}  ones its presence results in the appearance of a
magnetic flux. We have also learned how to move from one picture
to its dual. The above discussion naturally extends to any map $g$
from a two-dimensional manifold into the $SU(2)$ group. The
generalization to $SU(N)$ does involve some subtleties; following
\cite{BT}, we therefore briefly summarize the fundamental steps.
Given the WPL, patchwise we  can always write
 \begin{equation}
 \label{diag1}
 \Omega = h^{-1}_\alpha s h_{\alpha}
 \end{equation}
where $s$ is a diagonal matrix and $h_\alpha$ is map from the
patch $U_\alpha$ into $SU(N)$. In each patch, we can also
introduce another  matrix $g_\alpha$,
 \begin{equation}
 \label{functiong}
 g_\alpha=h^{-1}_\alpha \mu h_\alpha
 \end{equation}
where $\mu$ is a regular element in the Lie algebra of the maximal
torus, an integer linear combination of the diagonal generators of
$SU(N)$. One can show that $g$ is globally defined and thus drop
the index $\alpha$; since $\Omega$ is assumed to be regular and
since our $g$ commutes with $\Omega$, the obstruction to
diagonalizing $\Omega$ is also carried by $g$. Choose $\mu$ to be
one of the roots $\alpha^k$ of $SU(N)$. Then we can express $N-1$
integer-valued indices $n^k$ describing the aforementioned obstruction
in terms of $g$,
 \begin{equation}
 \label{eq37}
n^{k}(g,\mbox{\boldmath$A$}) = -\frac{1}{4\pi}\int {\rm Tr}
g[D(\mbox{\boldmath$A$},g),D(\mbox{\boldmath$A$},g)]
-\frac{1}{2\pi}
 \int {\rm Tr} g F_{A}\;\;,
 \end{equation}
where $D(\mbox{\boldmath$A$},g) = \mbox{\boldmath$A$} - h^{-1}dh$.
One might wonder why we introduced $g$, rather than working
directly with $\Omega$. Briefly, global diagonalizability of a
matrix is a property of the reference frame defined by its
eigenvectors; since we  consider regular maps, the explicit form
of its eigenvalues is irrelevant and only their non-degeneracy
matters. Therefore introducing $g$ is a simple tool for extracting
only the relevant information. This also explain why we were able
to give invariant meaning to the magnetic flux that emerged upon
diagonalization: we have singled out a set of intrinsic
vectors in the color space that
transform covariantly under gauge transformations, and used
them to bleach the color index, thus providing a (pseudo-)abelian
framework.\footnote{If we try to diagonalize $g$, and consequently
$\Omega$, it is not difficult to argue that, even in this more
general $SU(N)$ case, the components of the connection along the
Cartan subalgebra become non-trivial in the sense that they can
acquire non-vanishing ``$U(1)$" Chern classes -- one for each
diagonal generator of $SU(N)$. For a more detailed discussion of
this construction, see \cite{BT}.}

The appearance of the new topological index  and of the associated
abelian fluxes is an intriguing new feature, whose presence could
affect the Dirac determinant effective action, especially its
parity violating part. We have in mind analogous effects due to
magnetic fields in $QED_3$ or to instantons in $QCD_4$. To
quantify its repercussions, however, would require difficult
calculations we have not attempted.

\subsection{Examples}
We now proceed to illustrate the above general mechanism with two
explicit examples, $S^1\times S^2$ and $S^1\times T^2\simeq T^3$.
Consider first the following $SU(2)$ connection on ${\cal M}=S^1\times
S^2$,
\begin{equation}
\label{AA0A}
A_0(\theta,\phi)= \frac{2\pi \gamma}{\beta}\left(
\matrix{\cos\theta &\sin\theta e^{-i n \phi} \cr \sin\theta e^{i n
\phi} & -\cos\theta}\right),\ \ \ \ \  \mbox{\boldmath$A$}=0;
\end{equation}
it is globally defined when the angles ($\theta$,$\phi$)
span the unit sphere. Note that there is no  magnetic flux through
the sphere since the spatial components $\mbox{\boldmath$A$}$
vanish.\footnote{We also remark that the constant $\gamma$ cannot
be an integer, since then WPL becomes the identity, obviously not
a {\it regular} map (it commutes with any element of $SU(2)$).
Physically, $\gamma$ plays the role of a flat connection and
integer-valued $\gamma$ allow $A_0$ to be completely gauged away.
We also mention that making $\gamma$ angle-dependent does not
alter the results below. Nothing essential would change either had
we chosen $\mbox{\boldmath$A$}
=\mbox{\boldmath$A$}(t,\mbox{\boldmath$x$})$ instead of
$\mbox{\boldmath$A$}=0$. In fact, if
$\mbox{\boldmath$A$}(t,\mbox{\boldmath$x$})$ is a globally defined
connection on the sphere, its overall effect on the abelian gauge
field $(a_\theta, a_\phi)$ below is to add a topologically trivial
fluctuation, given by $Tr(A_0 \mbox{\boldmath$A$})$,  that would not alter the
value of the Chern class.}  What happens if we try to diagonalize
the WPL, $\exp(i \beta A_0)$,  associated with the above
(time-independent) $A_0$? The diagonalizing transformation
\begin{equation}
U_n(\theta,\phi)= \left( \matrix{ \cos(\theta/2) &  -e^{-i n
\phi}\sin(\theta/2)\cr \sin(\theta/2)e^{i n \phi} &
\cos(\theta/2)}\right) \label{treuno}
\end{equation}
is {\it not} globally defined on the sphere: for example,
$\phi$-dependence clearly remains at the  south pole,
$\theta=\pi$. Hence this gauge transformation is defined only in
the north pole chart of the sphere and another one, regular in the
south pole chart, must be constructed in order to achieve
consistent diagonalization. For the moment, let us work only
around the north pole; it will be clear how to extend  the final
result to the south pole region. Performing (\ref{treuno}) on
(\ref{AA0A}) yields $A_0=2\pi \gamma \sigma_3/\beta$, but also
introduces space components:
\begin{equation}
A^U_\theta=-\frac{i}{2}\left( \matrix{0 & e^{-i n\phi}\cr -e^{i
n\phi} &0}\right) \; , \ \ \ \ \  A^U_\phi=\frac{n}{2}
\left(\matrix{2 \sin^2(\theta/2)& -e^{ -i n\phi}\sin\theta\cr -
e^{i n\phi}\sin\theta & -2\sin^2(\theta/2)}\right) \; .
\label{tredue}
\end{equation}
We can now construct the apparently abelian field that keeps track
of the hidden original obstruction to diagonalizing  $A_0$, namely
\begin{equation}
a_\theta \equiv \frac{1}{2}{\rm Tr}(\sigma_3 A^U_\theta)=0 \; ; \
\ \ \ \ a_\phi \equiv \frac{1}{2}{\rm Tr}(\sigma_3 A_\phi^U)=n
\sin^2\frac{\theta}{2}= \frac{n}{2}(1-\cos\theta).
\end{equation}
We immediately recognize the usual monopole on the two-sphere, or
more precisely, its usual expression in the north pole chart. The
corresponding field strength is proportional to the volume form of
the sphere and therefore carries a non-vanishing Chern class with
value $n$. Analysis of  a gauge
transformation regular about the south pole yields the
expression for the same monopole in that chart. As expected from
general theory, the gauge transformation connecting the two
expression belongs to the maximal torus: it is proportional to
$\sigma_3$. For completeness, we write the gauge invariant form of
this index in terms of the above fields. If we call $g=U_n\sigma_3
U_n^{-1}/2=\beta\Omega/4\pi\gamma$, then in any gauge, the index
is
\begin{equation}
\label{cip12}
n=-\frac{1}{32\pi}\int_{S^2}{\rm Tr}(g[d_A g,d_A g]),
\end{equation}
which is the same as (\ref{genindex}) upon dropping the magnetic field term,
since $B$ vanishes in all gauges. Note incidentally that, being
gauge invariant, (\ref{cip12}) could appear in the effective action for
such backgrounds.

It is  instructive to consider an example defined on a
different 2-surface, the 2-torus $T^2$, so that now
${\cal M}=S^1\times T^2\simeq T^3$. The
spatial torus $T^2$ is parametrized by two flat periodic
coordinates $x_1\sim x_1+1$ and $x_2\sim x_2+1$. We start by
considering a transformation $V(x_1,x_2)$ of $SU(2)$ periodic up to an element
$\omega_i$ of the maximal torus; in particular we
choose
\begin{eqnarray}
\label{periodicity}
 &&V(x_1+1,x_2)=\omega_1 V(x_1,x_2)= \left(\matrix{e^{-2\pi
i x_2 }&0\cr0 & e^{2\pi i x_2} }\right)V(x_1,x_2)\\[.1in]
&&V(x_1,x_2+1)=\omega_2 V(x_1,x_2)=V(x_1,x_2),
\end{eqnarray}
where the explicit form of $\omega_i$ is the trivial embedding in
$SU(2)$ of the transition functions of the $U(1)$-instanton with
unit charge  on $T^2$. We can now construct a globally defined
({\it i.e.}, periodic) WPL as follows
\begin{eqnarray}
\Omega(x_1,x_2)&=&V^{-1}(x_1,x_2)\left(\matrix{\exp(i\phi(x_1,x_2))
&0\cr 0& \exp(-i\phi(x_1,x_2))}\right) V(x_1,x_2)=\nonumber\\
[.15in]
 &=&
\left(\matrix{|\alpha|^2 e^{i\phi}+\mid\beta\mid^2 e^{-i\phi}& 2 i
\bar\alpha\bar \beta \sin\phi\cr -2 i \alpha\beta\sin\phi &
e^{i\phi}\mid \beta\mid^2+\mid\alpha \mid^2 e^{-i\phi} }\right),
\end{eqnarray}
in terms of the parametrization
\begin{equation}
\label{parame1}
 V(x_1,x_2)=\left(\matrix{\alpha(x_1,x_2)&
\bar\beta(x_1,x_2)\cr -\beta(x_1,x_2)& \bar\alpha(x_1,x_2)}\right)
, \;\;\; |\alpha|^2+|\beta|^2=1  \; .
\end{equation}
The eigenvalues $\exp(\pm i\phi)$ of $\Omega$ are taken to be
periodic. The periodicity, and consequently the global nature of
$\Omega$, instead follows from the fact that the transformation
$\omega_i$ belongs to the maximal torus.

Now we show that attempting to diagonalize the above WPL gives
rise to a magnetic flux. The additional components along the
Cartan subalgebra are, as already explained in the previous
example,
\begin{equation}
A_i(x_1,x_2)=\frac{1}{2} {\rm Tr}[\sigma_3(\partial_i
V(x_1,x_2))V^{-1}(x_1,x_2)].
\end{equation}
They satisfy the following periodicity conditions
\begin{eqnarray}
&& A_1(x_1+1,x_2) = \frac{1}{2} {\rm Tr}[\sigma_3(\partial_1
V(x_1+1,x_2))V^{-1}(x_1+1,x_2)]=A_1(x_1,x_2),\\
&& A_2(x_1+1,x_2) = \frac{1}{2} {\rm Tr}[\sigma_3(\partial_2
V(x_1+1,x_2))V^{-1}(x_1+1,x_2)]  =  A_1(x_1,x_2)-2\pi, \\
&& \mbox{\boldmath $A$} (x_1 , x_2 +1 ) = \mbox{\boldmath $A$}
(x_1 , x_2 +1 ).
\end{eqnarray}
This is exactly the behavior of a (unit charge) instanton on
$T^2$. We thus can say equivalently that this field carries a
magnetic flux equal to one or, in the regular gauge, that WPL
defines a mapping with unit winding number. Finally, we must
demonstrate that a matrix $V(x_1,x_2)$ satisfying
(\ref{periodicity}) indeed exists; in the parametrization
(\ref{parame1}), $(\alpha , \beta )$ must obey
\begin{eqnarray}
\alpha(x_1+1,x_2)&\!\!\!\!=\exp(-2\pi i x_2) \alpha(x_1,x_2)\ \ \
\ \ \alpha(x_1,x_2+1)&\!\!\!\!=\alpha(x_1,x_2)\\
\beta(x_1+1,x_2)&\!\!\!\!\!\!=\exp(2\pi i x_2) \beta(x_1,x_2)\ \ \
\ \ \  \ \beta(x_1,x_2+1)&\!\!\!\!=\beta(x_1,x_2).
\end{eqnarray}
These periodicity requirements are solved in terms of
$\Theta-$functions, namely by
\begin{eqnarray}
\alpha(x_1,x_2)= \frac{1}{\mathcal{N}}\bar\Theta \left[{x_2 \atop
x_1}\right](0,i) \ \ \ \beta(x_1,x_2)=
\frac{1}{\mathcal{N}}\Theta\left[{x_2\atop{
x_1+\lambda}}\right](0,i)
\end{eqnarray}
where the normalizing factor $\mathcal{N}$ maintains the condition
$|\alpha|^2+|\beta|^2=1$ and $\lambda$ is any (non-integer)
constant. Our construction also suggests a simple and natural
representative LGT on $T^3$, namely
\begin{equation}
U_n(t,x_1,x_2)=V^{-1}(x_1,x_2) e^{i \pi n t\sigma_3/\beta}
V(x_1,x_2)e^{-i \pi n t\sigma_3/\beta} \; ;
\end{equation}
it is globally defined on $T^3$ and has winding number $n$. [A similar
construction in a $QCD_4$ context was examined in \cite{Wipf}.]

\section{The Fate of LGT}
\setcounter{equation}{0}
Throughout the literature \cite{DGS1,NS,Dunne,Salcedo,bdf,RSF,Das}
on effective nonabelian actions, a difficult question, as we have
already mentioned, has been the role and form of LGT, a name
indiscriminately attributed to two different classes of
transformations. The first is that (discussed in Sec.\ 2) of the
usual color LGT with nonvanishing $W_2(U)$. The second is that of
the ``abelian'' LGT, whose action, in the gauge where the WPL is
diagonal, is to shift the flat connection wrapping around the
Euclidean time direction by an integer-valued diagonal matrix. In
this gauge, it has the simple form
\begin{equation}
\label{hjk} u(t)=\exp\left(\frac{2\pi i t }{\beta} D\right)
\end{equation}
where $D$ is a linear integer combination of the diagonal
generators of $SU(N)$. In most explicit computations, the question
of the invariance of the finite temperature effective actions
under LGT  refers to this second class. These transformations, as
is manifest when written in the form (\ref{hjk}), can be unwrapped
smoothly, hence they
cannot be considered a subset of the actual  LGT.
Calling them ``LGT" at this level merely reflects
 their formal resemblance to the abelian ones. They are in
fact LGT for the subgroup $U(1)^{N-1}$, but they seemingly lose
this property when immersed in $SU(N)$: their winding number
$W_2(u)$, when computed from (\ref{hjk}), vanishes.  In the
following, we will show that they are nevertheless LGT  in a
deeper sense: The ``abelian'' LGT appearing in  the gauge where
the WPL is diagonal and a magnetic flux is present become, in the
dual picture where all the fields are made regular,
transformations which are genuinely large ($W_2\ne 0$). This
dualism between large and small transformations is  the
manifestation of a more general problem: the meaning and the
definition of LGT when we deal with non-trivial bundles, or more
generically, with bundles whose transition functions are not the
identity. In fact, the index $W_2(U)$ in the form (\ref{W3}) is
not sufficient to capture the winding of a transformation when we
use a non-global section of the gauge-bundle. For us, this problem
was not dramatic since we just considered $SU(N)$, which admit
only trivial bundles and thus we always had a global section from
which to check if a transformation is large. This problem becomes
unavoidable, however, for non-simply-connected group or dimensions
different from three; one such case was indeed briefly discussed
in Sec. 2.

It is convenient here to select the well-known (almost) temporal gauge
\begin{equation}
\label{A0dot} \dot A_0\equiv \partial_0 A_0=0\ \ \
\Leftrightarrow\ \ \ A_0(t,\mbox{\boldmath$x$})\equiv
\frac{2\pi}{\beta}\mathbb{
A}_0(\mbox{\boldmath$x$})\; ;
\end{equation}
periodicity prevents one from setting $A_0=0$, since this would
entail a trivial WPL. Investigating the role of LGT when we have
already picked a specific gauge might appear self-contradictory,
but our choice leaves a residual freedom that includes
representatives of large gauge transformations.
The fact that this gauge is attainable is shown for example in
\cite{000}. Here we briefly review the main steps of its
construction since they are necessary in understanding the
residual gauge group. Imposing the gauge (\ref{A0dot}) amounts to
finding periodic solutions of the linear differential equation
\begin{equation}
\label{coppa12}
\partial_t U(t,\mbox{\boldmath$x$}) +i A_0(t,\mbox{\boldmath$x$})
 U(t,\mbox{\boldmath$x$})= \frac{2\pi i}{\beta} U(t,\mbox{\boldmath$x$})
\mathbb{A}_0(\mbox{\boldmath$x$})\; , \;\;\;
U(\beta,\mbox{\boldmath$x$})=U(0,\mbox{\boldmath$x$}) \;.
\end{equation}
In terms of the new variable
$V(t,\mbox{\boldmath$x$})=\Omega^{-1}(t,\mbox{\boldmath$x$})
U(t,\mbox{\boldmath$x$})$, (\ref{coppa12})  reduces to the
homogeneous equation,
\begin{equation}
\label{coppa13}
\partial_t V(t,\mbox{\boldmath$x$})=\frac{2\pi i}{\beta}
V(t,\mbox{\boldmath$x$})
\mathbb{A}_0(\mbox{\boldmath$x$}),
\end{equation}
since $\Omega(t,\mbox{\boldmath$x$}) = P \exp (-i\int^t_0
dt^\prime A_0 (t^\prime , \mbox{\boldmath$x$}))$ obeys
$\partial_t\Omega(t, \mbox{\boldmath$x$})
+iA_0(t,\mbox{\boldmath$x$}) \Omega(t,\mbox{\boldmath$x$})= 0$.
Since $\mathbb{A}_0(\mbox{\boldmath$x$})$ is time-independent,
(\ref{coppa13}) can be promptly integrated
\begin{equation}
V(t,\mbox{\boldmath$x$})= \hat V(\mbox{\boldmath$x$}) \exp\left( \frac{2\pi i
t}{\beta}\mathbb{ A}_0(\mbox{\boldmath$x$})\right).
\end{equation}
Thus the general solution of (\ref{coppa12}) is
\begin{equation}
\label{coppa14}
U(t,\mbox{\boldmath$x$})=\Omega(t,\mbox{\boldmath$x$}) \hat
V(\mbox{\boldmath$x$}) \exp\left(\frac{2\pi i t}{\beta}\mathbb{A}_0(\mbox{\boldmath$x$})\right).
\end{equation}
Imposing  the periodicity condition
$U(\beta,\mbox{\boldmath$x$})=U(0,\mbox{\boldmath$x$})$ determines
the possible values $\mathbb{ A}_0(\mbox{\boldmath$x$})$ in terms of
WPL, according to
\begin{equation}
\label{coppa15} \hat V^{-1}(\mbox{\boldmath$x$})\Omega(\beta,
\mbox{\boldmath$x$}) \hat V(\mbox{\boldmath$x$}) =\exp\left( -2\pi i
\mathbb{A}_0(\mbox{\boldmath$x$})\right),
\end{equation}
{\it i.e.,} $\mathbb{ A}_0$ is essentially the  logarithm of WPL.
Whether this logarithm always defines a global quantity is not
obvious; however one can see easily that, at least for regular
maps, it does. [Even in the general case this logarithm can
actually be safely taken, but establishing that requires more
careful analysis.]

We now probe the residual gauge freedom remaining after the choice
(\ref{A0dot}); (\ref{coppa14}) immediately implies that any
transformation of the form
\begin{equation}
\label{coppa16} U(t,\mbox{\boldmath$x$})=\exp\left(-\frac{2\pi it}{\beta}
\mathbb{A}_0(\mbox{\boldmath$x$})\right) \hat V(\mbox{\boldmath$x$})
\exp\left(\frac{2\pi i t}{\beta}\mathbb{B}_0(\mbox{\boldmath$x$})\right),
\end{equation}
preserves  (\ref{A0dot}) if
\begin{equation}
\label{coppa17} \hat V^{-1}(\mbox{\boldmath$x$})\exp\left( -2\pi i
\mathbb{ A}_0(\mbox{\boldmath$x$})\right)\hat
V(\mbox{\boldmath$x$}) =\exp\left( -2\pi i \mathbb{
B}_0(\mbox{\boldmath$x$})\right),
\end{equation}
which is just (\ref{coppa15}), but only within the class of
gauge-equivalent fields that  respect (\ref{A0dot}). Let us change
variables from $\mathbb{B}_0(\mbox{\boldmath$x$})$ to
$\Lambda(\mbox{\boldmath$x$})$ by means of
\begin{equation}
\mathbb{B}_0(\mbox{\boldmath$x$})=\hat
V^{-1}(\mbox{\boldmath$x$})\left(\mathbb{A}_0(\mbox{\boldmath$x$})+
\Lambda(\mbox{\boldmath$x$})\right )\hat V(\mbox{\boldmath$x$}).
\end{equation}
This final shift to the variable $\Lambda$ then leads to the
simple but powerful statement
\begin{equation}
\exp(2\pi i\mathbb{A}_0(\mbox{\boldmath$x$}))=\exp\left(2\pi i
\left(\mathbb{A}_0(\mbox{\boldmath$x$})+
\Lambda(\mbox{\boldmath$x$}) \right)\right)\; .
\end{equation}
If $\exp(2\pi i \mathbb{A}_0(\mbox{\boldmath$x$}))$ is a regular
map, this implies, for all $\mbox{\boldmath $x$}$, the two
conditions
\begin{equation}
[\exp(2\pi i
\mathbb{A}_0(\mbox{\boldmath$x$})),\Lambda(\mbox{\boldmath$x$})]=0,\
\ \ \exp(2\pi i \Lambda (\mbox{\boldmath$x$}))=1 \; ,
\end{equation}
so the eigenvalues of $\Lambda$ are integers. We can thus conclude
that the most general transformation that preserves our gauge
choice is
\begin{equation}
U(t,\mbox{\boldmath$x$})=\exp\left(\frac{2\pi i}{\beta} t
\Lambda(\mbox{\boldmath$x$})\right) \hat V(\mbox{\boldmath$x$}).
\end{equation}
Now it is immediate to recognize a connection with the previous
section: $\Lambda(\mbox{\boldmath$x$})$ is  nothing but the
(properly normalized) auxiliary function $g(\mbox{\boldmath$x$})$
introduced in (\ref{functiong}). More generally it could be any
integer combination of the $g$ associated to the different roots.
Thus if $A_0(\mbox{\boldmath$x$})$ corresponds to a field whose
WPL is not smoothly diagonalizable, the function
$\Lambda(\mbox{\boldmath$x$})$ can be chosen to carry this
information, through having a non-trivial index.
As a consequence,
\begin{equation}
U(t,\mbox{\boldmath$x$})=\exp\left (\frac{2\pi i}{\beta}
t\Lambda(\mbox{\boldmath$x$})\right)
\end{equation}
preserves the gauge (\ref{A0dot}); it is both LGT and periodic.
The LGT property follows from the fact the
$\Lambda(\mbox{\boldmath$x$})$ is not smoothly diagonalizable. In
fact, recalling that $\Lambda(\mbox{\boldmath$x$})$ has been
identified with $g(\mbox{\boldmath$x$})=h^{-1} \alpha h$, we can
write \be \label{eq51}
\partial_t U= \frac{2\pi i}{\beta} U \Lambda \ \ , \ \ \ \ \ \ \
\partial_i U=[U, h^{-1} \partial_i h].
\ee
 In turn, the index of the transformation can be recast as
\be W_2(U)=\frac{i}{8\pi\beta} \int^{\beta}_0 dt \int_\Sigma  dx
\epsilon^{ij}{\rm Tr}(\Lambda [U^{-1} [U, h^{-1} \partial_i h],
U^{-1}[U, h^{-1} \partial_j h]])\; .
 \ee
 It is easy to see that the integrand in $W_2$ is the sum of two terms,
 \be
 X \equiv \e^{ij} \: Tr \: \Lambda [h^{-1} \pa_i h, h^{-1} \pa_j
 h]\; , \;\;\;
 Y \equiv \e^{ij} Tr \Lambda [h^{-1} \pa_i h, U^{-1} h^{-1} \pa_j h] \;
 ,
 \ee
 and that $Y$ is a total time derivative whose
 integral vanishes by periodicity, while $X$ is time-independent.
 Hence, it easily follows that
 \be
 W_2 (U) = 1/4\pi \int d^2x X = 2  n_2 (\L) \; .
 \ee
But $ n_2(\Lambda)$ is precisely the value, at $\mbox{\boldmath$A$}=0$,
of the index
defined in (\ref{eq37}).  This finally proves that the residual group
for non-diagonalizable WPL still contains (the ``true") LGT. Of
course, $\Lambda$ is neither fixed nor unique: there are in
general $N-1$ linearly independent matrices $\Lambda^i$ that
commute with $A_0$. Therefore we can construct in principle a wide
array of transformations by taking integer linear combinations of
$\Lambda^i$. Unfortunately, this space does not contain all the
large gauge transformations, but just those whose index is an even
multiple of the flux  associated to the torus bundle. However,
their action on $A_0$ is simply the shift
\begin{equation}
A_0\to A_0+2\pi \Lambda \;.
\end{equation}
When we diagonalize the WPL (and consequently $A_0$), $\Lambda$
also diagonalizes since it commutes with $A_0$. In terms of these
diagonal forms $(a_0)$, the above
equation reduces to
\begin{equation}
a_0\to a_0+2\pi D ,
\end{equation}
where $D$ is an integer linear combination of the diagonal
generators of $SU(N)$.  These transformations can be interpreted to
act on the constant part of  the eigenvalues of $a_0$; but this is
exactly the result of a transformation of the form (\ref{hjk}).

Summarizing, when we gauge-diagonalize WPL (which does not alter
our gauge condition $\dot A_0=0$), the surviving LGT are further
reduced to the simpler form (\ref{hjk}). This is, in a certain
sense, expected because we limit our gauge functions to those of
the torus bundle in the process and the surviving LGT share this
fate: They become LGT of the torus bundle gauge group
$U(1)^{N-1}$. When the fluxes vanish, obviously no large
transformation survives the chosen gauge fixing. In this limit,
the flat connection becomes essentially irrelevant and non-trivial
interplays between topological quantities in the magnetic sector
are absent. Only the (topologically dull) electric sector remains.

\section{Vanishing $\mbox{\boldmath$E$}$}
\setcounter{equation}{0}
The final example in Sec.\ 3  illustrated the seemingly
paradoxical, extreme situation of starting with no magnetic field
and  ending up with a field on the maximal torus that actually
carries a flux. In the recent literature \cite{Salcedo,RSF,Das} on
explicit calculations of effective actions, the opposite extreme
has also been considered, where instead, $\mbox{\boldmath$E$}$
vanished. In the discussions of the general structure of these
latter configurations, a certain amount of confusion has arisen,
whose origin lies in non-gauge invariant descriptions of the
fluxes discussed here.  Hence we next review the general form of
such fluxes.

Since the vanishing of $\mbox{\boldmath$E$}$ is a gauge invariant
property, we simplify our analysis by choosing the gauge of Sec.\
4, and from now assume that
$A_0(t,\mbox{\boldmath$x$})=\mathbb{A}_0(\mbox{\boldmath$x$})$;
the latter is not arbitrary however, since $\Omega=\displaystyle{
\exp\left(2\pi i \mathbb{A}_0(\mbox{\boldmath$x$})\right) }$,
and $\Omega$ is constrained to be covariantly constant \cite{GPY,WPL}
\begin{equation}
D_i \Omega\equiv\partial_i\Omega+[A_i(0,\mbox{\boldmath$x$}),\Omega]=0 \;.
\end{equation}
Its gauge invariant content can be understood by taking the
derivative of ${\rm Tr}(\Omega^n)$:
\begin{equation}
\partial_i ({\rm Tr}(\Omega^n))=n {\rm Tr}(\Omega^{n-1}\partial_i\Omega)=
n {\rm Tr}(\Omega^{n-1}D_i\Omega)=0,
\end{equation}
{\it i.e.}, the trace of any power of $\Omega$ is independent of
$\mbox{\boldmath$x$}$, and its eigenvalues are all constant.
Similarly, the Bianchi identities imply that the magnetic field
$B\equiv F_{12}$ is covariantly conserved,
\begin{equation}
D_t B(t,\mbox{\boldmath$x$})=\partial_t B(t,\mbox{\boldmath$x$})+i
[\mathbb{A}_0 (\mbox{\boldmath$x$}), B(t,\mbox{\boldmath$x$})]=0
\; ,
\end{equation}
or equivalently,
\begin{equation}
\partial_t \left [ \exp\left(-\frac{2\pi i t}
{\beta}\mathbb{A}_0(\mbox{\boldmath$x$})\right)
B(t,\mbox{\boldmath$x$})\exp\left(\frac{2\pi i
t}{\beta}\mathbb{A}_0(\mbox{\boldmath$x$})\right) \right ]=0 \; .
\end{equation}
When integrated between $(0,t)$, this yields
\begin{equation}
\label{but} B(t,\mbox{\boldmath$x$})= \exp\left(\frac{2\pi i
t}{\beta}\mathbb{A}_0(\mbox{\boldmath$x$})\right)
B(0,\mbox{\boldmath$x$})\exp\left(-\frac{2\pi i
t}{\beta}\mathbb{A}_0(\mbox{\boldmath$x$})\right) \; .
\end{equation}
This does not yet imply  that $B(t,\mbox{\boldmath$x$})$ can
be made time-independent, because the WPL does not define a
legitimate, periodic, gauge transformation\footnote{We allow only
periodic transformations since for the moment we want to have
globally defined potentials, namely ones that do not differ by a
gauge transformation when we change charts on $S^1$.} still. The
periodicity condition $B(0,\mbox{\boldmath$x$})=
B(\beta,\mbox{\boldmath$x$})$ does require the WPL to commute with
$B(0,\mbox{\boldmath$x$})$,
\begin{equation}
\label{578}
\left [B(0,\mbox{\boldmath$x$}),
\exp\left(2\pi i \mathbb{A}_0(\mbox{\boldmath$x$})\right)
\right]=0.
\end{equation}
In the following we limit ourselves to {\it regular} WPL (recall
again that a regular map means, for $SU(N)$, that all its
eigenvalues are different); most of the result will not depend on
this technical assumption; we shall comment further on this point.
Then (\ref{578}) immediately entails that
$[\mathbb{A}_0, B(0, \mbox{\boldmath$x$})]=0$, which, combined
with (\ref{but}), is equivalent to the time-independence of $B$.

So far, we have seen that one we can always achieve a gauge where
$A_0$ and $B$ are both time-independent and commute.  The
following steps become much easier if we further diagonalize $A_0$
and $B$, as can be always arranged because the residual gauge
freedom contains all time-independent transformations. The price
is that the components of $\mbox{\boldmath$A$}$ along the Cartan
subalgebra might live  on a non-trivial torus bundle, but we have
learned how to deal  with this difficulty in Sec.\ 3. Thus, from
now on we have, in our ``torus gauge",
\begin{equation}
\mathbb{A}_0 (\mbox{\boldmath$x$})= {a}_0^i  H_i\
\ \ {\rm and} \ \ \ B (\mbox{\boldmath$x$})= {\cal
B}^i(\mbox{\boldmath$x$}) H_i.
\end{equation}
where the $H_i$ generate  the Cartan subalgebra. The factor
$a_0^i$ are constant, being linear combinations of the eigenvalues
of the WPL. The vanishing of $\mbox{\boldmath$E$}$ allows us to
make both $B$ and $\mbox{\boldmath$A$}$ time-independent. In fact,
in our diagonal gauge, this condition ($\mbox{\boldmath$E$}\equiv
\partial_t \mbox{\boldmath$A$}+[{a}_0,\mbox{\boldmath$A$}]=0$)
can be easily integrated between $(0,t)$ to yield
\begin{eqnarray}
\label{raff}
\mbox{\boldmath$A$}(t,\mbox{\boldmath$x$})=\exp\left(-\frac{2\pi i t}{\beta}
a_0\right)\mbox{\boldmath$A$}(0,\mbox{\boldmath$x$})
\exp\left(\frac{2\pi i t}{\beta}
{a}_0\right).
\end{eqnarray}
This has the same structure as (\ref{but}) and, by the same
arguments, leads to the analogous conclusion,
\begin{equation}
[a_0,\mbox{\boldmath$A$}(0,\mbox{\boldmath$x$})]=0,
\end{equation}
and consequently to time-independence,
$\mbox{\boldmath$A$}(t,\mbox{\boldmath$x$})=
\mbox{\boldmath$A$}(0,\mbox{\boldmath$x$})$.

Summarizing, for vanishing electric field, if the WPL is a regular
map, we can always choose a gauge where all connections are
time-independent, $A_0$ is also space-independent and finally
$[A_0,\mbox{\boldmath$A$}]=0$. This configuration can carry a
magnetic flux.\footnote{It has been claimed \cite{Das} that there
exist configurations with vanishing $\mbox{\boldmath$E$}$ that are
not gauge equivalent to those obtained above. In particular, it
was stated that there are, in this class, connections that are
genuinely time-{\it dependent}. However, two points were
overlooked: First, the possibility of diagonalizing $A_0$ through
residual gauge transformations was not exploited. Second,  only
globally defined gauge transformations were allowed, a requirement
we saw to be too restrictive once non-trivial torus bundles were
admitted.} In particular our result shows that if
$\mbox{\boldmath$E$}=0$, a magnetic flux must be present for there
at all to be a nontrivial effective action in the PV
sector: otherwise the candidate CS action, $\int Tr(a_0 B)$,
vanishes.

We now briefly comment on non-regular maps. In general, this
question is very involved, as one must know the general structure
of the sub-manifold where $\Omega$ has degenerate eigenvalues. For
us, $\mbox{\boldmath$E$}=0$ provides a substantial simplification;
the eigenvalues do not depend on $\mbox{\boldmath$x$}$, and
therefore the presence of degenerate ones is also
$\mbox{\boldmath$x$}$-independent. This protects us from
pathologies such as permutations of the eigenvalues arising from
their flow through the spatial surface, so our results persist
also in this context.

\section{Summary}
\setcounter{equation}{0}
We have studied the interplay between color (already present at
T=0) and thermal LGT, and the consequences of the ensuing
topological complications.  The central quantity that carries the
large information is the WPL, generalizing the flat connection of
the simple abelian U(1). We provided the relevant mathematical
framework for keeping track of the mapping involved in this
``double LGT" world. We then exhibited explicit non-singular LGT,
studied their effects on WPL, and stressed the novel index that
measures topological obstruction to its global diagonalization. We
also linked the index to extension of the usual global WPL
diagonalization process with a wider class of transformations that
are not globally defined, but require use of local charts. Our
explicit constructions, besides the mentioned LGT example,
included the gauge potentials in temporal gauge for configurations
with nonvanishing index.  Then, for vanishing
$\mbox{\boldmath$E$}$ (that we do not believe to be an essential
restriction), we showed precisely how ``magnetism without
magnetism" arises, {\it i.e.}, how transforming configurations
with vanishing B gives rise to gauge invariant magnetic flux
properties in the Cartan subalgebra directions. In the process, we
corrected some misconceptions in the recent literature.

The clarifications we hope to have provided here should be thought
of as an entry to a number of issues we have left untouched.  To
mention a few, the effect of an index on the parity-violating
parts of the effective action/Dirac determinant might be quite
extensive; perhaps it could be probed using some simple but
indicative special configurations.  Any fallout to nonperturbative
D=4 effects would of course be particularly important, but may be
more remote in view of the pivotal role of odd dimension in our
considerations.  Finally, the ``kinematics" we have attempted to
sort out should be of use in a more explicit analysis of QCD$_3$
dynamics, perhaps without recourse to particular field
configurations.

\noindent{\bf Acknowledgements}:  This work was supported by the
National Science Foundation under grant PHY99-73935; D.S.\ was
supported by TMR grant FMRX CT96--0045.
D.S. and L.G. were
supported by the Bruno Rossi program during the final phase of
this work, begun in the previous millennium; both thank MIT for
its warm hospitality. We thank C. Isham for
useful correspondence.

  \end{document}